\documentclass[twoside,twocolumn,9pt]{article}
\usepackage{extsizes}
\usepackage[super,sort&compress,comma]{natbib} 
\usepackage[version=3]{mhchem}
\usepackage[left=1.5cm, right=1.5cm, top=1.785cm, bottom=2.0cm]{geometry}
\usepackage{balance}
\usepackage{mathptmx}
\usepackage{sectsty}
\usepackage{graphicx} 
\usepackage{lastpage}
\usepackage[format=plain,justification=justified,singlelinecheck=false,font={stretch=1.125,small,sf},labelfont=bf,labelsep=space]{caption}
\usepackage{float}
\usepackage{fancyhdr}
\usepackage{fnpos}
\usepackage[english]{babel}
\addto{\captionsenglish}{%
  \renewcommand{\refname}{Notes and references}
}
\usepackage{array}
\usepackage{droidsans}
\usepackage{charter}
\usepackage[T1]{fontenc}
\usepackage[usenames,dvipsnames]{xcolor}
\usepackage{setspace}
\usepackage[compact]{titlesec}
\usepackage{hyperref}

\usepackage{epstopdf}

\definecolor{cream}{RGB}{222,217,201}

\begin{document}

\pagestyle{fancy}
\thispagestyle{plain}
\fancypagestyle{plain}{
\renewcommand{\headrulewidth}{0pt}
}

\makeFNbottom
\makeatletter
\renewcommand\LARGE{\@setfontsize\LARGE{15pt}{17}}
\renewcommand\Large{\@setfontsize\Large{12pt}{14}}
\renewcommand\large{\@setfontsize\large{10pt}{12}}
\renewcommand\footnotesize{\@setfontsize\footnotesize{7pt}{10}}
\makeatother

\renewcommand{\thefootnote}{\fnsymbol{footnote}}
\renewcommand\footnoterule{\vspace*{1pt}%
\color{cream}\hrule width 3.5in height 0.4pt \color{black}\vspace*{5pt}} 
\setcounter{secnumdepth}{5}

\makeatletter 
\renewcommand\@biblabel[1]{#1}            
\renewcommand\@makefntext[1]%
{\noindent\makebox[0pt][r]{\@thefnmark\,}#1}
\makeatother 
\renewcommand{\figurename}{\small{Fig.}~}
\sectionfont{\sffamily\Large}
\subsectionfont{\normalsize}
\subsubsectionfont{\bf}
\setstretch{1.125} 
\setlength{\skip\footins}{0.8cm}
\setlength{\footnotesep}{0.25cm}
\setlength{\jot}{10pt}
\titlespacing*{\section}{0pt}{4pt}{4pt}
\titlespacing*{\subsection}{0pt}{15pt}{1pt}

\fancyfoot{}
\fancyfoot[LO,RE]{\vspace{-7.1pt}\includegraphics[height=9pt]{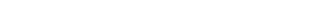}}
\fancyfoot[CO]{\vspace{-7.1pt}\hspace{13.2cm}\includegraphics{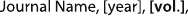}}
\fancyfoot[CE]{\vspace{-7.2pt}\hspace{-14.2cm}\includegraphics{head_foot/RF}}
\fancyfoot[RO]{\footnotesize{\sffamily{1--\pageref{LastPage} ~\textbar  \hspace{2pt}\thepage}}}
\fancyfoot[LE]{\footnotesize{\sffamily{\thepage~\textbar\hspace{3.45cm} 1--\pageref{LastPage}}}}
\fancyhead{}
\renewcommand{\headrulewidth}{0pt} 
\renewcommand{\footrulewidth}{0pt}
\setlength{\arrayrulewidth}{1pt}
\setlength{\columnsep}{6.5mm}
\setlength\bibsep{1pt}

\makeatletter 
\newlength{\figrulesep} 
\setlength{\figrulesep}{0.5\textfloatsep} 

\newcommand{\topfigrule}{\vspace*{-1pt}%
\noindent{\color{cream}\rule[-\figrulesep]{\columnwidth}{1.5pt}} }

\newcommand{\botfigrule}{\vspace*{-2pt}%
\noindent{\color{cream}\rule[\figrulesep]{\columnwidth}{1.5pt}} }

\newcommand{\dblfigrule}{\vspace*{-1pt}%
\noindent{\color{cream}\rule[-\figrulesep]{\textwidth}{1.5pt}} }

\makeatother

\twocolumn[
  \begin{@twocolumnfalse}
{\includegraphics[height=30pt]{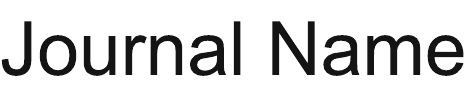}\hfill\raisebox{0pt}[0pt][0pt]{\includegraphics[height=55pt]{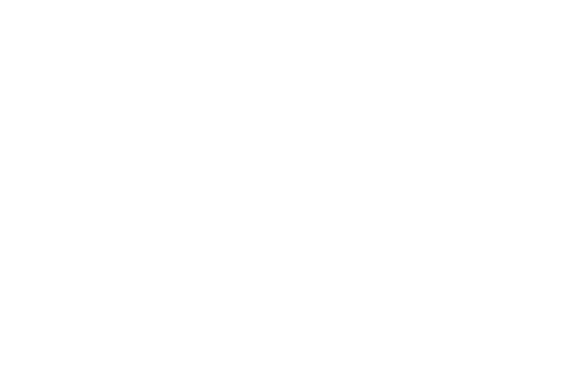}}\\[1ex]
\includegraphics[width=18.5cm]{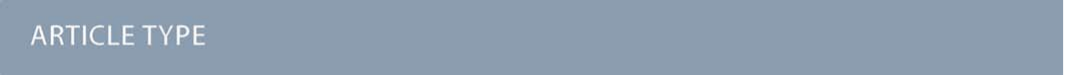}}\par
\vspace{1em}
\sffamily
\begin{tabular}{m{4.5cm} p{13.5cm} }

\includegraphics{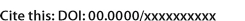} & \noindent\LARGE{\textbf{Transfer learning of GW-Bethe--Salpeter Equation excitation energies$^\dag$}} \\
\vspace{0.3cm} & \vspace{0.3cm} \\

 & \noindent\large{Dario Baum,\textit{$^{a}$} Arno Förster,\textit{$^{a}$} and Lucas Visscher$^{\ast}$\textit{$^{a}$}} \\

\includegraphics{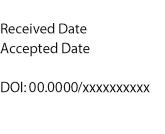} & \noindent\normalsize{A persistent challenge in machine learning for electronic-structure calculations is the sharp imbalance between abundant low-fidelity data like DFT or TDDFT results and the scarcity of high-fidelity data like many-body perturbation theory labels. We show that transfer learning provides an effective route to bridge this gap: graph neural networks pretrained on DFT and TDDFT properties can be finetuned with limited qs$GW$ and qs$GW$-BSE data to yield accurate predictions of quasiparticle and excitation energies. Assessing both full-model and readout-only finetuning across chemically diverse test sets, we find that pretraining improves accuracy, reduces reliance on costly qs$GW$ data, and mitigates large predictive outliers even for molecules larger or chemically distinct from those seen during finetuning. Our results demonstrate that multi-fidelity transfer learning can substantially extend the reach of many-body-level predictions across chemical space.} \\

\end{tabular}

 \end{@twocolumnfalse} \vspace{0.6cm}

  ]

\renewcommand*\rmdefault{bch}\normalfont\upshape
\rmfamily
\section*{}
\vspace{-1cm}


\footnotetext{\textit{$^{a}$~Department of Chemistry and Pharmaceutical Sciences, Vrije Universiteit Amsterdam, De Boelelaan 1108, 1081 HZ Amsterdam, The Netherlands; E-mail: l.visscher@vu.nl}}


\footnotetext{\dag~Electronic supplementary information (ESI) available: Details on test set compositions, model hyperparameters, training parameters and additional analyses like error bars and learning curves. See DOI: XY}



\section{Introduction}



Accurate prediction of properties of materials and molecules underpins advances across chemical physics \cite{ChemPhys_Example_1, ChemPhys_Example_2, ChemPhys_Example_3, ChemPhys_Example_4}, materials science \cite{Aakash_Example1, Aakash_Example_2, Aakash_Database}, and molecular design \cite{MolDesign_Example_1, MolDesign_Example_2} for applications like optoelectronic materials \cite{SOTA_ML-GW_SolarMat, MaxMalte_OptoMat, Gardner_Spectroscopy} and catalysis \cite{Margraf_BEEF, Margraf_EnBarriers}. Especially excited-state properties such as excitation energies are fundamental for studying processes like photosynthesis \cite{Photosynth_Bsp_1, Photosynth_Bsp_2} and photovoltaic energy conversion \cite{EnConversion_Bsp_1, EnConversion_Bsp_2, SOTA_ML-GW_SolarMat}. For such purposes, computational methods are frequently employed because experiments on electronically excited states, such as determining excitation energies or characterizing short-lived molecular intermediates, are often prohibitively complex \cite{ExpsHard_1, ExpsHard_2, ExpsHard_3}. High‐level wavefunction methods, most notably equation-of-motion coupled cluster (EOM-CC) \cite{EOM-CC_1, EOM-CC_2}, provide excellent descriptions of charged and neutral excitations and systematically converge to the full configuration interaction limit for weakly correlated excited states \cite{accuracy_EOM_CC_1, accuracy_EOM_CC_2, accuracy_EOM_CC_3}. Yet even truncated variants such as EOM-CCSD (single and double excitations) and EOM-CCSDT (single, double and triple excitations) exhibit steep computational scaling, restricting their routine use to small molecules. Conversely, and time-dependent (TD)\cite{TDDFT_1, TDDFT_2} density functional theory (DFT)\cite{DFT_1, DFT_2} is dramatically cheaper but also significantly less accurate\cite{TDDFT_quanitatively_inaccurate}. Many-body perturbation theory (MBPT)\cite{GWA_4, GWA_5} offers a more favourable accuracy–cost balance. In particular, the GW approximation\cite{GWA_1, GWA_3, photoemission_1, photoemission_2, Marie2024b} to the electronic self-energy and its combination with the Bethe–Salpeter equation (GW-BSE) \cite{BSE_1, BSE_2, GWA_2, IPA_1} yields quasiparticle (QP) energies, and optical excitation energies with accuracy rivaling high-level wavefunction benchmarks \cite{GW_accuracte_1, mol_benchmarks_14, GW100_2, GW_accuracte_2, GW_accuracte_5, GW_accuracte_6, BSE_accuracte_2, jacquemin2017bethe, knysh2024reference}. Particularly the quasi-particle self-consistent variant (qs$GW$)\cite{Faleev2004, qsGW_1, qsGW_2} of the GW approximation frequently yields excellent results\cite{GW_accuracte_3, GW_accuracte_4, duchemin2025joint, BSE_accurate_1} and at the same time eliminates the dependence on the underlying mean-field reference \cite{LowOrderScaling_qsGW}. Although being almost as efficient as TDDFT\cite{BSE_accurate_1}, qsGW and qsGW-BSE remain computationally demanding, limiting throughput and the breadth of chemical space accessible.

Machine learning (ML) models promise to bridge this gap by providing accuracy of high-fidelity electronic-structure data at a fraction of the computational cost. Early work used kernel methods to predict GW QP energies, Green’s functions, and molecular orbital levels \cite{SOTA_GW-ML_Baumeier, SOTA_ML-GW_Venturella, SOTA_ML-GW_Stuke}. More recently, graph neural networks (GNNs) have demonstrated accurate and transferable predictions of total energies, orbital energies, and excited-state properties, enabled by sufficiently large datasets of molecular properties \cite{QM9_GW_SOTA, QM9_1, QM9_2}. Models such as MACE \cite{MACE_1, MACE_2, MACE_3}, SchNet \cite{SchNet_1, SchNet_2}, DimeNet++ \cite{DimeNetPP} and OptiMate \cite{OptiMate_1, OptiMate_2} have been successfully trained to predict molecular and material properties such as evGW \cite{evGW} QP energies and gaps of organic molecules \cite{SOTA_GW-ML_Fediai} or the dielectric function of semiconductors and insulators \cite{OptiMate_1, OptiMate_2}.

A major bottleneck, however, is the scarcity of high-fidelity data, e.g. wavefunction methods, MBPT, or experimental labels, relative to the abundance of lower-fidelity data, e.g. DFT and TDDFT. Large resources such as OMol25 \cite{OMol25}, QCML \cite{QCML} and QCDGE \cite{QCDGE} provide DFT-labels for tens of millions of molecules or TDDFT-level labels for hundreds of thousands of molecules. Opposed to that, existing datasets at the wavefunction or MBPT level like OE62 \cite{OE62}, QM9GWBSE \cite{QM9GWBSE} and GDB-9-Ex\_EOMCCSD \cite{TDDFT_and_EOMCCSD_data} comprise orders of magnitude fewer samples. This disparity raises a key question: how can GNN models exploit abundant low-fidelity data while achieving high-fidelity accuracy?

Recent studies have begun to address this multi-fidelity challenge through different pretraining and finetuning strategies \cite{MultifidelityTL, MaxMalte_Finetuning, TL_Gardner, Gardner_Distillation} and targeting, for example, pretraining on DFT or semiempirical data and finetuning, for instance, on Coupled Cluster (CC) \cite{CCSD(T)_1, CCSD(T)_2, TL_CC, TL_Kästner, TL_MultiFidelity_Gardner} or Random-Phase-Approximation (RPA) labels \cite{RPA, TL_MaxMalte, TL_Margraf}. Collectively, these results indicate that representations learned on low-fidelity data transfer remarkably well to high-fidelity tasks, potentially reducing the need for expensive reference calculations at the high-fidelity level.

\begin{figure*}
 \centering
 \includegraphics[height=7cm]{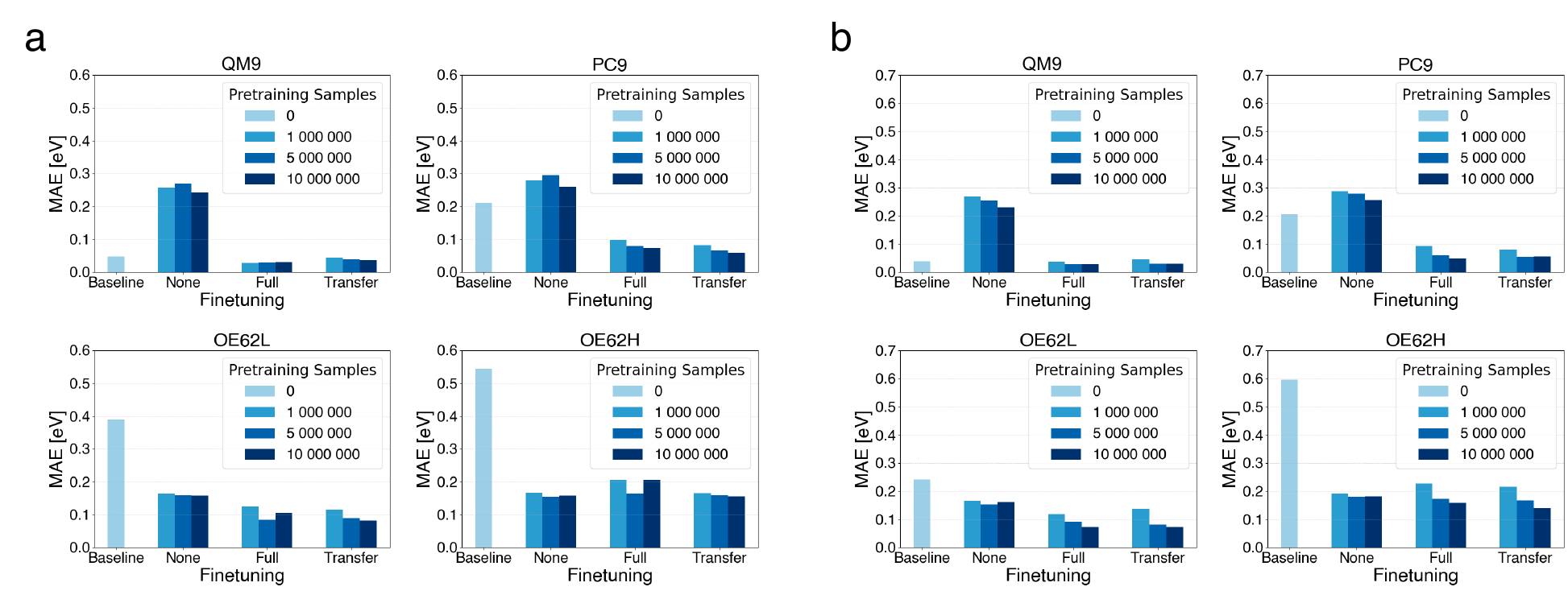}
 \caption{MAE of qs$GW$ QP HOMO predictions from small (a) and large (b) models pretrained on different numbers of DFT samples.}
 \label{fig: mae homo}
\end{figure*}

Contributing to these efforts, we investigate whether such multi-fidelity learning can accelerate the prediction of qs$GW$ and qs$GW$-BSE properties. 
We pretrain on DFT molecular orbital (MO) energies and TDDFT excitation energies and subsequently finetune on qs$GW$ and qs$GW$-BSE labels, respectively. In contrast to prior neural network models for $GW$ and $GW$-BSE which are typically trained from scratch \cite{SOTA_ML-GW_SolarMat, SOTA_GW-ML_Fediai, MBFormer}, employ $\Delta$-learning relative to DFT \cite{DeltaLearning, SOTA_GW-ML_Fediai}, or use descriptors from electronic structure calculations \cite{SOTA_ML-GW_SolarMat, MBFormer}, we leverage the multi-fidelity paradigm and provide end-to-end predictions directly from molecular structures to high-level excited-state observables. Thus, by using lower-fidelity data, we demonstrate how to achieve $GW$-BSE accuracy at ML computational cost, since no electronic structure calculation is needed for predictions in that way. We further demonstrate that a model pretrained on DFT and TDDFT data reduces the amount of expensive qs$GW$ and qs$GW$-BSE data needed for finetuning with no loss of accuracy. Together, these results establish multi-fidelity learning as a promising path toward data-efficient surrogate models that provide MBPT accuracy at ML speed for a diverse chemical space and thus enable rapid screening of excited states properties for large sets of diverse molecules, which would not be feasible with traditional MBPT methods.


\section{Methods}

\subsection{Data preparation}

We pretrain models on two types of quantum-chemical data. For MO energies, we use up to ten million neutral molecules from the OMol25 dataset with highest occupied molecular orbital (HOMO) energies, lowest unoccupied molecular orbital (LUMO) energies and HOMO-LUMO gaps at the $\omega$B97M-V/def2-TZVPD \cite{omegaB97M-V, VV10, def2} level of theory, which should give reasonably close estimates of the first ionization potential (IP), the electron affinity (EA), and the fundamental gap (IP-EA) respectively. To assess the effect of data scale, we also considered one- and five-million-molecule subsets. For excitation-energy pretraining, we used TDDFT excitation energies from the QCDGE dataset at the $\omega$B97X-D/6-31G(d) \cite{omegaB97X-D, 6-31Gd_1, 6-31Gd_2} level of theory. Finetuning was performed on qs$GW$ QP energies and qs$GW$-BSE excitation energies, both taken from the QM9GWBSE dataset.

As is standard for training of neural networks, we split QM9GWBSE into training, validation, and test data. The test data serves as the first test set to which we refer to as “QM9” for brevity.
Next to that, we test predictions on a subset of the PC9 \cite{PC9} dataset, which matches the QM9 element space (H, C, N, O, F) and molecular size (up to 29 atoms). For the remaining two test sets, we sampled molecules from OE62 that obey either (i) the QM9 element restrictions but allow up to 48 atoms, probing extrapolation to larger systems and referred to as “OE62L”, or (ii) QM9-sized molecules containing up to three heteroatoms not present in the finetuning data, probing generalization across chemical space and referred to as “OE62H”. This means that we test our models on four different datasets in total. All test-set labels were recomputed with the same qs$GW$ and qs$GW$-BSE settings used for the QM9GWBSE dataset. SMILES-based deduplication was applied across all datasets to prevent any overlap between pretraining, finetuning, and test molecules. Complete dataset specifications and filtering procedures are provided in the Supplementary Information (SI).

\subsection{Machine learning}
\begin{figure*}[hbt]
 \centering
 \includegraphics[height=6.75cm]{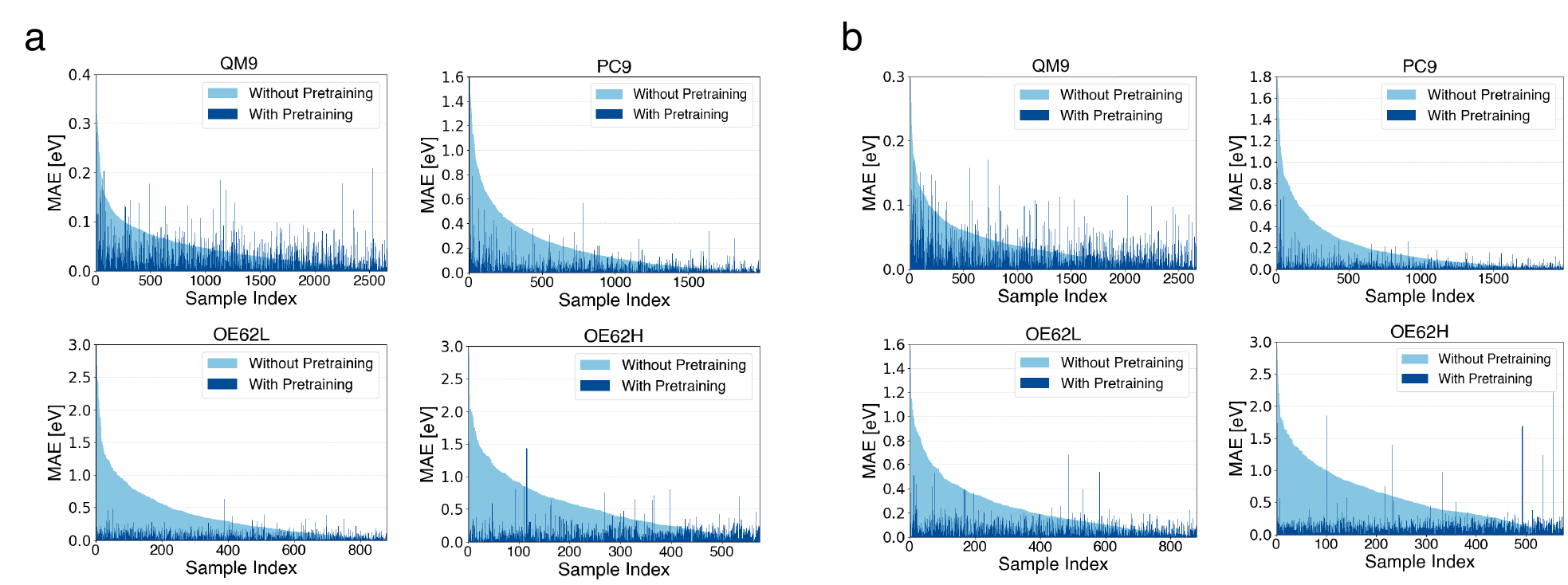}
 \caption{Per-sample absolute errors of qs$GW$ QP HOMO energy predictions from small (a) and large (b) models with and without pretraining.}
 \label{fig: improvement per sample homo}
\end{figure*}

We use the ViSNet architecture \cite{ViSNet}, which incorporates up to four-body interactions through treatment of bond lengths, bond angles, dihedral angles, and improper dihedrals, while maintaining computational efficiency which is crucial for large-scale pretraining. To assess the effect of model size, two model capacities are examined: a small variant with $8.9 \times 10^5$ parameters and a large variant with $2.5 \times 10^6$ parameters. Hyperparameters were selected to balance training efficiency, stability, and the accuracy of baseline (non-pretrained) models. To isolate the effect of pretraining, we maintained consistent hyperparameters between pretrained and non-pretrained runs, modifying them only when required to ensure stable convergence of training and validation losses.
We consider two finetuning strategies. In the “Full” approach, all model weights are updated during finetuning. In the “Transfer” approach, only weights in the readout layers following message-passing layers are updated, enabling reuse of latent representations learned from lower-fidelity data. Such transfer-learning strategies have proven effective for multi-fidelity molecular property prediction \cite{MultifidelityTL, TL_Maurer} and motivate their use here. Detailed hyperparameters and details on the training procedures are provided in the SI.

\section{Results}

We investigate whether a multi-fidelity learning paradigm can improve the accuracy of GNN-based predictions of qs$GW$ quasiparticle (QP) energies and qs$GW$–BSE excitation energies. Our analysis begins with qs$GW$ QP energies. We start by pretraining models on DFT MO energies and subsequently finetune them on qs$GW$ QP energies. Note that the DFT pretraining data is on the range-separated hybrid DFT level which is a reasonable approximation to corresponding QP energies \cite{gerber2005hybrid}. We then evaluate 1) whether pretraining reduces prediction errors on the test sets relative to training from scratch, thereby indicating improved generalization 2) whether pretraining enables the use of smaller finetuning datasets without sacrificing accuracy 3) whether model accuracies improve with pretraining on different but related properties, e.g. pretraining on HOMO energies and finetuning on HOMO-LUMO gaps. Finally, we pretrain on TDDFT excitation energies, then finetune on qs$GW$-BSE excitation energies and test, 4), whether analogous trends as in 1)-3) arise for this combination of pretraining and finetuning target. All results are compared to models initialized with random weights and trained solely and directly on the target property.

\begin{figure*}[hbt]
 \centering
 \includegraphics[height=4.75cm]{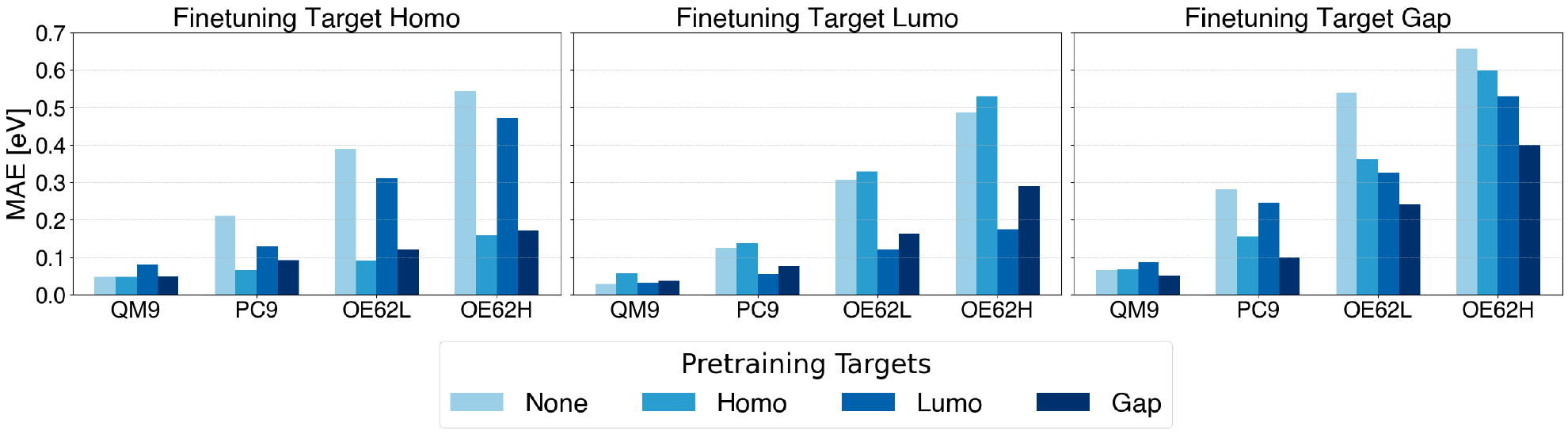}
 \caption{MAE of QP HOMO energy, QP LUMO energy and QP gap predictions with pretraining on different QP energy targets.}
 \label{fig: pretraining finetuning alignment}
\end{figure*}

\subsection{Does pretraining improve generalization?}

We first assess how model accuracy depends on the presence and extent of pretraining, model capacity, and finetuning strategy. Fig.~\ref{fig: mae homo} compares mean absolute errors (MAEs) across all test sets for the small and large models as a function of pretraining-set size, with 0 corresponding to baselines trained from scratch. We also report purely pretrained models (“None”), without any finetuning, to isolate the effect of pretraining alone.

Models undergoing both pretraining and finetuning consistently achieve the lowest MAEs across all test sets. The effect is particularly pronounced for PC9, OE62L and OE62H, where reductions of up to two-thirds are observed, for instance, in the large models on OE62H. Notably, on OE62L and OE62H the purely pretrained models perform nearly as well as their finetuned counterparts, indicating that for chemically challenging cases the models rely strongly on knowledge obtained during pretraining. Nonetheless, the lowest MAEs on every set are achieved when pretraining is followed by finetuning.

Increasing model capacity generally yields modest accuracy gains, especially for the OE62-based test sets. Because the larger model increases the dimensionality of the feature channels, it reduces information bottlenecks in message-passing and mitigates oversquashing \cite{OversquashingOversmoothing} which is especially relevant for long-range graph interactions and molecular structures with more diverse elements. However, for OE62H, both baseline models perform poorly, with almost no improvement from increased capacity. In contrast, pretraining enables both small and large models to make accurate predictions despite the presence of elements unseen during finetuning on the target property.

Transfer learning performs at least as well as, and frequently better than, full finetuning for any pretraining-set size. This indicates that, after pretraining on DFT level data, only a small subset of parameters requires updating to achieve optimal accuracy on qs$GW$ level predictions. For both model sizes, approximately 10\% of the number of parameters are finetuned. This observation aligns with reports of effective transfer learning in molecular-property prediction, for instance, in drug discovery tasks \cite{MultifidelityTL}.

To further elucidate the source of these improvements, we examine per-sample error changes. Fig.~\ref{fig: improvement per sample homo} compares absolute errors for models with and without pretraining, using the small transfer-learning model with 5\,000\,000 pretraining samples and the large model with 10\,000\,000 pretraining samples. In both plots, samples are sorted in descending order by their error without pretraining. Errors after pretraining are shown in the same order such that the $x$-axis represents the sample index.

Across all test sets, the largest improvements occur for samples with the highest baseline errors, while slight increases in error appear for samples that were already predicted accurately. As shown in Fig.~\ref{fig: improvement per sample homo}, the substantial reduction in errors for the most challenging samples far outweighs the modest increases among the easiest cases, explaining the strong MAE reductions on PC9, OE62L, and OE62H. The comparatively smaller improvement on QM9 reflects the absence of large outliers for that test set. The same qualitative trends are observed for the large model. Analogous analyses for QP LUMO energies and HOMO–LUMO gaps are provided in the SI and exhibit similar qualitative behavior.

Considering the trade-off between accuracy and computational cost, small transfer-learning models with 5\,000\,000 pretraining samples represent an effective compromise. This combination of models and pretraining set are therefore used as the default configuration in the following unless stated otherwise.

\subsection{Does pretraining reduce data demand in finetuning?}

\begin{figure}[hbt]
 \centering
 \includegraphics[height=7cm]{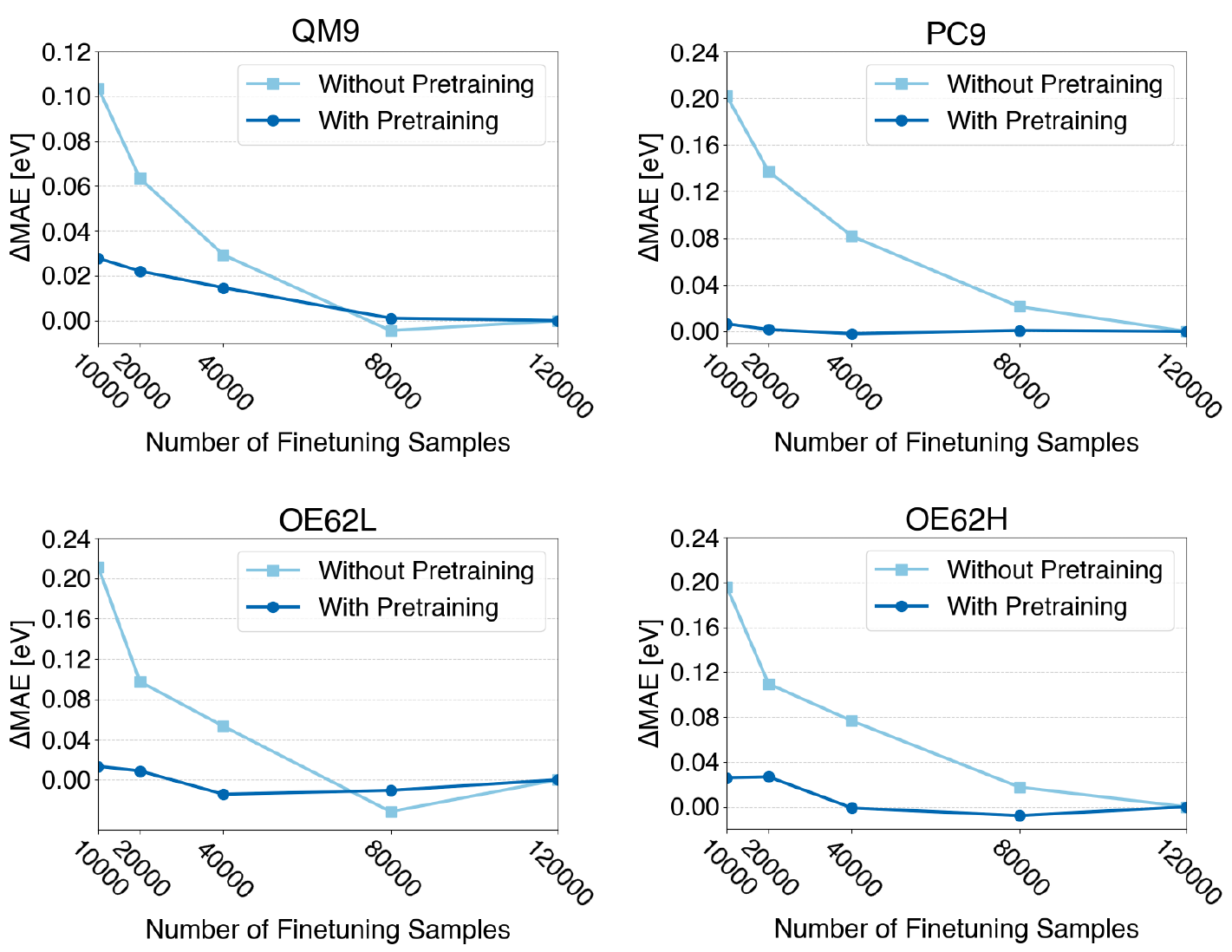}
 \caption{MAEs of QP HOMO energy predictions after finetuning on different numbers of samples with and without prior pretraining normalized to the respective MAE when finetuning on the full finetuning-set (120\,000 samples).}
 \label{fig: learning curves homo}
\end{figure}

Earlier work has shown that transfer learning between low- and high-fidelity data can reduce the amount of data required for finetuning \cite{MultifidelityTL, TL_Kästner}. Here, we assess whether this holds for qs$GW$ QP energies when pretraining on DFT MO energies. To this end, we construct finetuning sets of 10\,000, 20\,000, 40\,000, 80\,000 and, as before, the full set of 120\,000 samples. Each finetuning subset is obtained by independent sampling from the full QM9GWBSE dataset rather than by incremental augmentation, ensuring maximal randomization.

Models trained with and without pretraining are evaluated on all four test sets, and the resulting learning curves for QP HOMO energies are shown in Fig. \ref{fig: learning curves homo}. Corresponding curves for QP LUMO energies and gaps are provided in the SI. Each curve is shifted by its final error (i.e., the error at 120\,000 samples) so that the $y$-axis reflects the deviation from the presumed minimum MAE. For every finetuning-set size, three independently sampled subsets are generated, and the reported metrics are averages over these three runs, reducing the influence of small-sample effects.

Despite averaging, models trained from scratch still show a more irregular convergence, especially for OE62L, whereas the corresponding curves for pretrained models converge noticeably smoother. This is expected: without pretraining, each finetuning sample carries greater weight, and the removal of a single informative sample can produce abrupt changes in MAE. Pretraining effectively increases the total amount of information available, thereby mitigating such fluctuations.

Most importantly, Fig. \ref{fig: learning curves homo} shows that, across all test sets, pretrained models reach convergence with 40\,000 finetuning samples, and often with only 20\,000 samples, even for the most demanding cases. Thus, at most one-third of the qs$GW$ data required for training from scratch is sufficient to achieve comparable accuracy which effectively reducing the computation cost of prodcuing high-level training data. The same trends are observed for QP LUMO energies and QP gaps (see SI).

\subsection{Does pretraining transfer across properties?}

Beyond reduction of test errors and qs$GW$ data requirements, it is desirable for foundation models pretrained on one property to be reusable for finetuning on related targets. Such cross-property transfer would obviate repeated costly pretraining. To assess its feasibility, we examine all combinations of pretraining and finetuning on QP HOMO energies, QP LUMO energies, and QP gaps, and compare their performance with baselines trained directly on the target property. The resulting MAEs are shown in Fig. \ref{fig: pretraining finetuning alignment}.

In nearly all cases, pretraining on any of the three properties lowers the MAE relative to the baseline. As expected, the largest improvements occur when pretraining and finetuning targets coincide. However, substantial gains also arise when the two targets are information-theoretically linked: QP gap models benefit from HOMO or LUMO pretraining, and conversely, HOMO and LUMO models benefit from gap pretraining. Because these properties are inherently related, the gap being the LUMO–HOMO difference, pretraining exposes the model to patterns directly relevant to the downstream task, yielding a more favorable initialization than random weights. In contrast, when pretraining and finetuning targets share little underlying information, the benefits vanish and can even reverse, known as negative transfer \cite{NegTransfer_1, NegTransfer_2}, as reported for supervised pretraining in molecular representation learning \cite{PretrainForMolRepr}. For example, pretraining on QP HOMO energies and finetuning on QP LUMO energies increases MAEs across all test sets. 

Conclusively, cross-property transfer can yield sizable error reductions even approaching the gains of perfectly aligned pretraining. Nonetheless, realizing the full benefit of pretraining still requires that both targets coincide.

\subsection{Does pretraining and finetuning carry over to other excited-state targets?}

\begin{figure}[hbt]
 \centering
 \includegraphics[height=5cm]{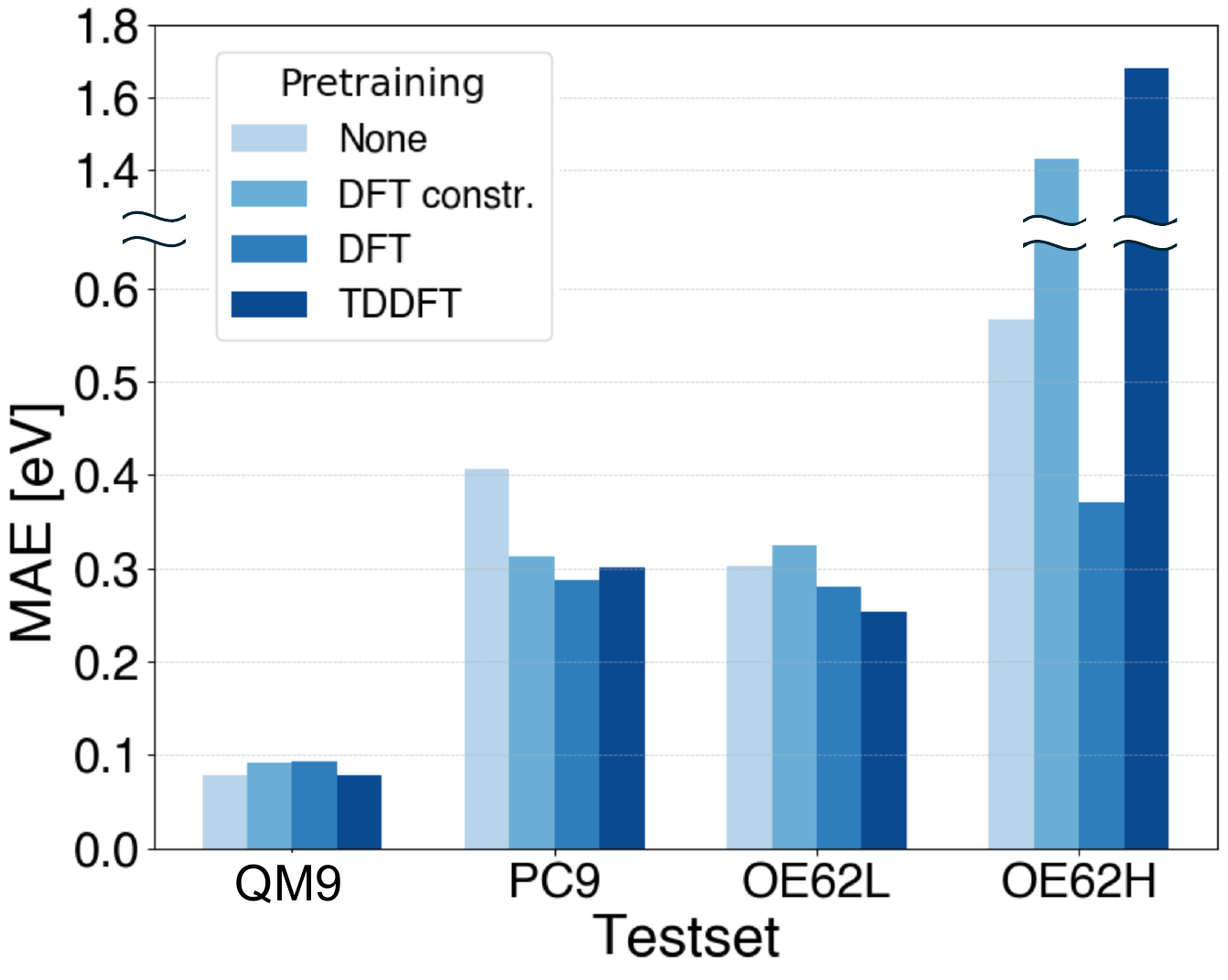}
 \caption{MAE of qs$GW$-BSE excitation energy predictions after pretraining on DFT and TDDFT data.}
 \label{fig: BSE errors}
\end{figure}

Finally, we assess whether our multi-fidelity strategy extends to qs$GW$-BSE excitation energies. First, we test whether finetuning on qs$GW$-BSE benefits from pretraining on TDDFT excitation energies. Because both methods describe neutral excitations, we expect reasonable alignment between the two targets. In all experiments, we use the lowest excitation energy per molecule. For comparison, we also evaluate models pretrained on DFT MO energies, hypothesizing that TDDFT should provide a more suitable pretraining signal than DFT. For DFT pretraining, we construct a 500\,000-sample set of DFT gaps from our OMol25 subset restricted to molecules containing H, C, N, O, and F and with at most 32 atoms, mirroring the element and size limits of the QCDGE dataset used for TDDFT pretraining. This isolates the effect of the pretraining target from the effect of dataset size and diversity. We additionally pretrain a second model on the full, unconstrained 5\,000\,000-sample DFT dataset used earlier to probe whether increased size and chemical diversity can compensate for a less well-aligned pretraining target. The resulting MAEs are shown in Fig. \ref{fig: BSE errors} where the constrained DFT set is denoted “DFT constr.”. Note that the $y$-axis is split because the OE62H errors are reported on a different scale.

\begin{figure}[hbt]
 \centering
 \includegraphics[height=7cm]{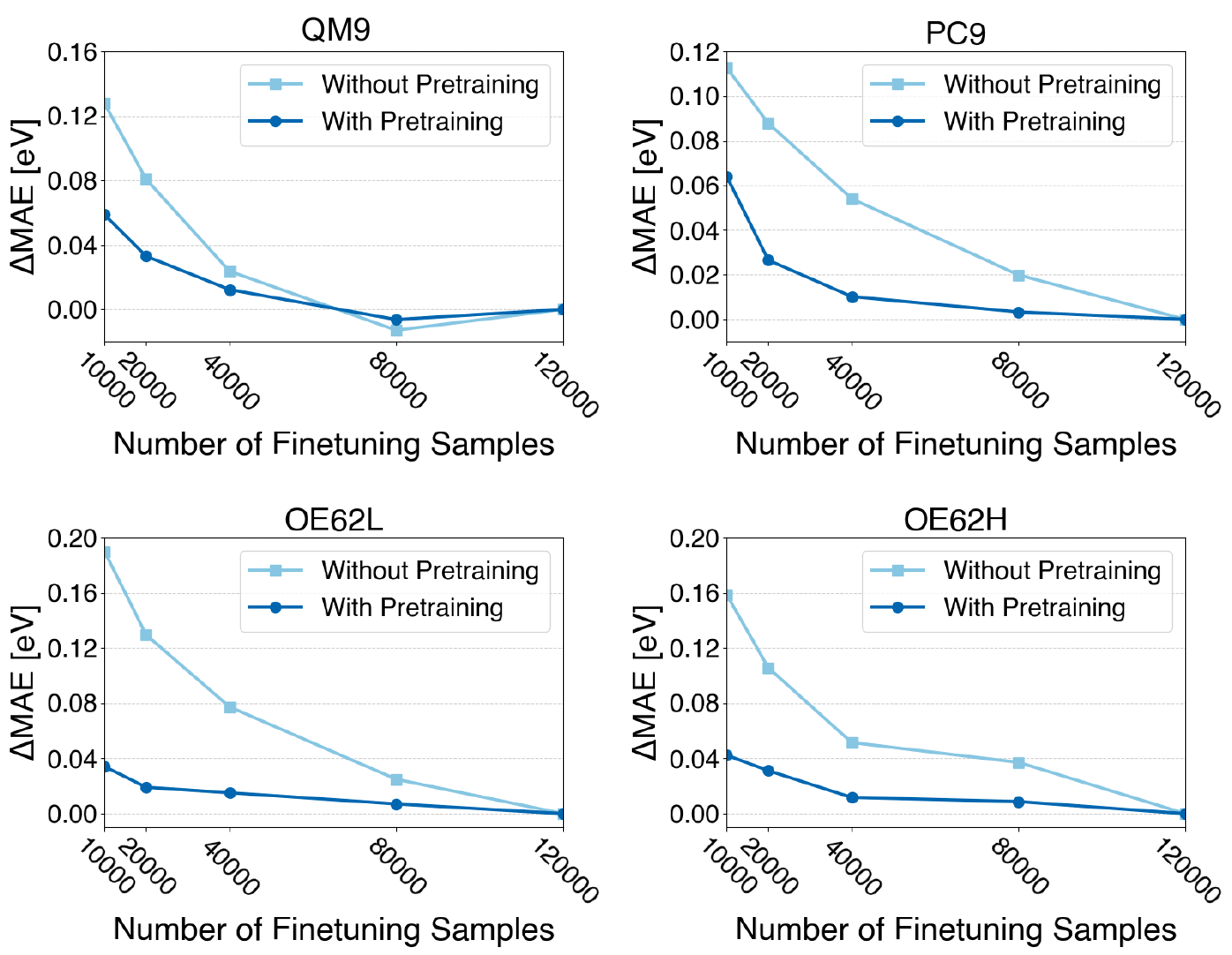}
 \caption{MAEs of qs$GW$-BSE excitation energy predictions after finetuning on different numbers of samples with and without prior pretraining normalized to the respective MAE when finetuning on the full finetuning-set (120\,000 samples).}
 \label{fig: learning curves bse}
\end{figure}

On QM9, baseline errors are already low and finetuning has only minor impact as observed before for QP energies. On PC9 and OE62L, TDDFT pretraining clearly improves MAEs, albeit less strongly than the transfer from DFT to qs$GW$ observed earlier. DFT pretraining on the constrained DFT set yields only marginal gains on OE62L and slightly worsens performance on PC9. The reason for that could be the misalignment between DFT gaps and qs$GW$-BSE excitations and is therefore in line with our previous findings about negative transfer due to insufficient alignment.

For OE62H, both TDDFT pretraining and DFT pretraining with a constrained dataset significantly increase the MAE. This is plausible, as both pretraining datasets span only a narrow element distribution, whereas OE62H contains a much more diverse range of elements. In contrast, DFT pretraining on the large, unconstrained dataset, despite its weaker target alignment, markedly reduces MAEs on OE62H presumably due to its broader coverage of chemical space. Similarly, although constrained DFT pretraining slightly increases the MAE on OE62L, the unconstrained model offers modest improvement. Notably, TDDFT pretraining still outperforms both DFT-based approaches on OE62L, underscoring its stronger alignment with qs$GW$-BSE. These trends suggest that larger and more chemically diverse TDDFT datasets could further lower MAEs across PC9, OE62L, and especially OE62H, potentially matching the gains observed with unconstrained DFT pretraining. Such datasets, however, are not currently available.

Second, we examine whether pretraining reduces the amount of costly qs$GW$–BSE data required for finetuning. Following the procedure used for QP energies, we construct finetuning sets of 10\,000, 20\,000, 40\,000, 80\,000, and the full 120\,000 samples. Again, each of those subset is drawn independently from the QM9GWBSE dataset. As before, models with and without pretraining are evaluated on all four test sets. For each test set, we apply the model variant that, in the previous analysis (Fig. \ref{fig: BSE errors}), achieved the lowest MAE when pretrained on the respective target property (DFT or TDDFT gap). Specifically, models pretrained on DFT gaps are used for PC9 and OE62H, while those pretrained on TDDFT gaps are used for QM9 and OE62L. The resulting learning curves are shown in Fig. \ref{fig: learning curves bse}. Each curve is shifted by its final error (i.e., the MAE obtained with 120\,000 finetuning samples), such that the $y$-axis reflects the deviation from that final error on the full finetuning set. 

As in the QP energy analysis, pretraining clearly accelerates convergence of the test set errors, most notably for PC9, OE62L, and OE62H. Quantitatively, the curves indicate that approximately 40\,000 samples, about one-third of the full qs$GW$–BSE finetuning set, are sufficient to reach convergence across all test sets. This mirrors the behavior observed for qs$GW$ QP energies and again demonstrates that pretraining can markedly reduce the amount of high-level data required to achieve target accuracy.

In summary, the multi-fidelity paradigm extends naturally to qs$GW$-BSE excitation energies, reducing test errors and reducing the computational cost of producing high-level labels for training. However, our findings also suggest that realizing the full benefits requires large pretraining datasets that are both well aligned with the target property and sufficiently broad in chemical diversity.
\section{Conclusions}

In this work, we investigated transfer learning from lower-fidelity DFT and TDDFT data to higher-fidelity qsGW and qsGW-BSE targets. Models were pretrained on DFT molecular orbital energies or TDDFT excitation energies and subsequently finetuned on qsGW quasiparticle energies or qsGW-BSE excitation energies, either by updating all weights or by restricting training to the readout layers following message-passing layers. The impact of these strategies was assessed across four test sets, including one containing molecules larger than those used during finetuning on the target property and another containing heteratoms also unseen during finetuning.

Our results show that pretraining on DFT- and TDDFT-level data provides a more favorable initialization of model weights than standard random initialization for learning qs$GW$ and qs$GW$-BSE properties. Notably, accurate models can be obtained by leveraging only a modest subset of the abundant DFT data and by training only a fraction of the model parameters, thereby reducing cost for both data-generation and training. We observe lower test errors, driven in particular by a reduction of large outliers observed without pretraining. On top of that, we demonstrate decreased requirement for expensive qs$GW$ and qs$GW$-BSE data. Specifically around 20\,000 instead of 120\,000 finetuning samples for qs$GW$ and around 40\,000 instead of 120\,000 finetuning samples for qs$GW$-BSE suffice to converge the errors on our most challenging test sets. We also show evidence of the transferability of foundation models across distinct QP properties. This presents a promising opportunity to “recycle” existing foundation models rather than generating new datasets and retraining models from scratch for each target property. Thus, with our proposed strategy we lower the expected amount of data needed for both pretraining and finetuning.

Overall, this study demonstrates that data-efficient GNN models capable of end-to-end prediction at the level of MBPT can be realized, with generalization extending into regions of chemical space not encountered during finetuning. At the same time, our findings underscore key challenges, including the need for sufficiently large and chemically diverse low-fidelity datasets. We show that the careful selection of pretraining targets of adequate level of theory that sufficiently align with finetuning targets are crucial to fully unlock the benefits of pretraining.

\section*{Author contributions}

D.B. conceptualized the project, conducted the calculations and analyzed the data. A.F. and L.V. provided guidance and supervision throughout the project. L.V. acquired funding that supported this work. D.B wrote the manuscript which was reviewed by  A.F. and L.V..

\section*{Conflicts of interest}

There are no conflicts to declare.

\section*{Data availability}

Code, test data and model checkpoints for this work are publicly available at \url{https://github.com/daoiradrio/ViSNetGWBSE}.

\section*{Acknowledgements}

We acknowledge the use of supercomputer facilities at SURFsara sponsored by NWO Physical Sciences, with financial support from The Netherlands Organization for Scientific Research (NWO). LV and DB acknowledge funding from Microsoft Research. AF acknowledges funding through a VENI grant from NWO under grant agreement VI.Veni.232.013.



\balance

\renewcommand\refname{References}

\bibliography{rsc} 
\bibliographystyle{rsc} 
\end{document}